\documentclass[pre,aps,showpacs,floatfix,nofootinbib,showkeys,preprint]{revtex4}

\usepackage{graphicx}
\usepackage{amsmath}
\usepackage[all]{xy}
\usepackage{amsfonts,amssymb}
\usepackage{comment}
\usepackage{multirow}

\begin{document}

\title{Traffic by multiple species of molecular motors}

\author{Yan Chai}
\author{Stefan Klumpp}
\author{Melanie J.I. M{\"u}ller}
\author{Reinhard Lipowsky}
\affiliation{
Max Planck Institute of Colloids and Interfaces,\\
Science Park Golm, 14424  Potsdam, Germany\\}


\begin{abstract}

We study the traffic of two types of molecular motors using the two-species asymmetric simple exclusion process (ASEP) with periodic boundary conditions and with attachment and detachment of particles. We determine characteristic properties such as motor densities and currents by simulations and analytical calculations. For motors with different unbinding probabilities, mean field theory gives the correct bound density and total current of the motors, as shown by numerical simulations. For motors differing in their stepping probabilities, the particle-hole symmetry of the current-density relationship is broken and mean field theory fails drastically. The total motor current exhibits exponential finite-size scaling, which we use to extrapolate the total current to the thermodynamic limit. Finally, we also study the motion of a single motor in the background of many non-moving motors.



\end{abstract}


\maketitle


\section{Introduction}

Intracellular transport by molecular motor proteins is essential for many cellular functions such as shuttling vesicles or organelles to their proper destinations \cite{Howard_2001, Schliwa_2003}. In eukaryotic cells,  polar cytoskeletal filaments serve as tracks for three families of cytoskeletal motors: kinesins and dyneins that move along microtubules, and myosins that walk along actin filaments \cite{Howard_2001, Schliwa_2003}.

The movements of a molecular motor along its filamentous track can be described as a biased random walk. However, after a certain number of steps the motor unbinds from the track  because the thermal noise of the environment overcomes the finite motor--filament binding energy. The motor then diffuses in the surrounding fluid as a Brownian particle until it comes close to the track to rebind to it. On large time scales, motors therefore perform peculiar 'motor walks' that consist of alternating periods of directed motion or biased random walks on the tracks and undirected diffusive motion in the fluid environment \cite{Lipowsky_Klumpp_2005}.

Intracellular transport is typically achieved by several types of motors rather than just by one type \cite{Hirokawa_1998, Vale_2003}. Different types of motors may either move independently along the same filament or work together to power bi-directional transport of a cargo that carries two types of motors \cite{Mueller_Lipowsky2008}. Here we consider the first case, where different species of motors move on the same filament, and interact only via their mutual exclusion. These motors species might move into different directions, or they might move into the same direction but have different velocities or different affinities to the filament. We are interested in the traffic behavior of the latter systems.

A simple model for motor traffic is based on the asymmetric simple exclusion process (ASEP) \cite{Schuetz_2001}. In the ASEP, particles that interact through mutual  exclusion from lattice sites perform biased random walks along a one dimensional lattice. This paradigmatic model for non-equilibrium transport exhibits interesting phenomena such as boundary induced phase transitions and traffic jams \cite{Schuetz_2001}. To describe the traffic of motor proteins, ASEP-like models have to be supplemented with the dynamics of binding to and unbinding from the track to account for the finite run length of the motors \cite{Lipowsky_Nieuwenhuizen_2001,Klumpp_Lipowsky_2003,Parmeggiani_Frey_2003,Popkov_Schuetz_2003,Evans_Santen_2003}.

ASEPs with a single species of particles can be generalized to models with multiple species \cite{Derrida_1998, Blythe_Evans_2007}. Even without attachment and detachment of particles, multi-species ASEPs exhibit a variety of cooperative phenomena, such as spontaneous symmetry breaking \cite{Evans_Mukamel_1995a} and phase separation \cite{Evans_Mukamel_1998, Arndt_Rittenberg_1998a}.

To model cellular traffic with different types of motors, one has to consider multi-species ASEPs with attachment and detachment of particles. Apart from investigations on conditions for the exact solvability \cite{Jafarpour_Khaki_2007, Arita_2008}, most work on multi-species ASEPs with detachment and attachment has focused on the traffic of two species of particles moving into opposite directions \cite{Evans_Mukamel_2002, Klumpp_Lipowsky_2004, Levine_Willmann_2004, Jafarpour_Ghavami_2007a, Ebbinghaus_Santen_2009}. These models exhibit continuous phase transitions \cite{Evans_Mukamel_2002, Klumpp_Lipowsky_2004, Jafarpour_Ghavami_2007a}, hysteresis of the total current \cite{Klumpp_Lipowsky_2004}, traffic jams \cite{Ebbinghaus_Santen_2009}, spontaneous symmetry breaking and localized shocks \cite{Levine_Willmann_2004}. Here we focus on the case where two species of particles move in the same direction, but do so either with different velocities or with different binding affinities to the filament. A few such systems have been studied experimentally in vitro, in particular the traffic of mixtures of fast-moving and slow- or non-moving kinesin motors \cite{Seitz_Surrey_2006}.

The paper is organized as follows: In section \ref{Theoretical model for multi-species motor traffic} we will present the theoretical model for multi-species motor traffic. We will focus on the situation that two motor species differ in only one parameter. In section \ref{Two species of motors with different unbinding probabilities} we will discuss the traffic of two types of motors which differ only in their unbinding parameters. In section \ref{Two species of motors with different stepping probabilities} we will discuss the traffic of two species with different stepping parameters, specifically one species with zero velocity. We will end with a summary and discussion in section \ref{Summary and discussion}.


\section{Theoretical model for multi-species motor traffic}
\label{Theoretical model for multi-species motor traffic}

To investigate cellular transport or \textit{in vitro} experiments with multiple species of molecular motors, we study multi-species ASEP models with attachment and detachment of particles. As mentioned in the introduction section, motors perform random walks, which are biased towards the same direction for all motor species, along a cytoskeletal filament. Such filaments are polymers which have periodic motor binding sites with repeat distance $\ell$, and motors walk along the filament with steps of the same size $\ell$. For microtubules, the lattice constant is $\ell=8$\ nm. We therefore represent the filament by a one-dimensional lattice of $L$ binding sites with spacing $\ell$. Per unit time $\tau$, a bound motor of type $i$ on the filament makes a forward step with probability $\alpha_{\rm i}$ if the target site is free, unbinds with probability $\epsilon_{\rm i}$ and remains at the same site otherwise \cite{Lipowsky_Nieuwenhuizen_2001}. Unbound motors are treated as a motor reservoir \cite{Klumpp_Lipowsky_2003, Parmeggiani_Frey_2003} with unbound density $\rho_{\rm ub,i}$ for the $i$-th type of motors. An unbound motor binds to the filament with the sticking probability $\pi_{\rm ad,i}$ when the binding site is empty. The treatment of unbound motors as a reservoir is slightly different from that in a previous lattice model for motor traffic in Ref.\ \cite{Lipowsky_Nieuwenhuizen_2001} where the diffusion of unbound motors has been treated explicitly as Brownian movement. The simplification to treat the unbound motors as a reservoir is appropriate if the density of unbound motors is homogeneous, as is the case for sufficiently fast diffusion and large motor numbers in the bulk solution. This is typically the case in \textit{in vitro} experiments, where the buffer is an aqueous solution with motor concentrations in the nano- or micro-molar range. The typical length of a microtubule is of the order of $10\mu \rm{m}$ or even larger, which is one order of magnitude larger than the typical motor run length, which is of the order of $1\mu \rm{m}$ for kinesins. Therefore, we use periodic boundary conditions
throughout this paper. This simplification is appropriate as most motors do not feel the ends of the filament, so that effects of the filament ends can be neglected. \footnote{Note also that we do not scale the binding and unbinding probabilities of motors with the system size $L$. The latter scaling, which has been used in Ref.\ \cite{Parmeggiani_Frey_2003}, leads to a motor run length that is comparable to the filament length $L$ in the limit of large $L$.} In our simulations, the length $L$ of the filaments is chosen sufficiently large so that finite-size effects can be ignored, as we show explicitly in Sec.\ \ref{Two species of motors with different stepping probabilities} below. \footnote{In most of our simulations, the filament length $L$ was chosen to be $L=200$ corresponding to about $2\mu \rm{m}$. In the presence of slow motors, the actual run length of the fast motors is strongly reduced and small compared to $L=200$. In the absence of slow motors, as in Sec.\ \ref{Two species of motors with different unbinding probabilities}, the length of the filament is irrelevant as well since our system then exhibits mean-field behavior for all $L$.}

\begin{figure}[h]
\centering
\includegraphics[width=7.5cm,clip]{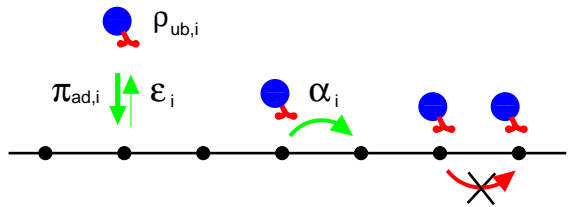}
\caption{(Color) The model for multi-species motor traffic. Per unit time $\tau$, bound motors of species $i$ hop one step forward along the filament with probability $\alpha_{\rm i}$ and detach from the filament with probability $\epsilon_{\rm i}$, while unbound motors with density $\rho_{\rm ub,i}$ attach to an empty site with probability $\pi_{\rm ad,i}$. Exclusion interactions between motors prohibit the movement of motors if the target site is occupied by another motor. The system is on a ring of $L$ sites with lattice constant $\ell$.}
\label{fig_multispecies_model}
\end{figure}

To sum up, as shown in Fig.\ \ref{fig_multispecies_model}, during each time step $\tau$ one of the $L$ sites is randomly chosen and updated according to the following stochastic dynamical rules:
\begin{eqnarray}
\text{Stepping:        }  i\, 0 \rightarrow 0\, i &&  \text{with probability  } \alpha_{\rm i} ,\label{model_stepping}\\
\text{Detachment:  }  i \rightarrow 0 && \text{with probability  } \epsilon_{\rm i} ,\label{model_detachment}\\
\text{Attachment:  }  0 \rightarrow i && \text{with probability  } \rho_{\rm ub,i}\pi_{\rm ad,i} \label{model_attachment},
\end{eqnarray}
where $0$ denotes a vacancy site and $i$ represents a site occupied by a motor particle of species $i$. All probabilities for other transitions are zero.

We performed Monte Carlo simulations on the the one-dimensional periodic lattice with $L$ sites. In the simulations, we control the number of motors in the system by changing the value of the unbound motor densities. This mimics \textit{in vitro} experiments where the motor concentration in the buffer is an easily accessible control parameter. Cells can also control the motor concentration in the cytoplasm by regulating gene expression. Once the unbound motor densities have been settled, we use a random-sequential update. Each simulation step, which corresponds to a unit of the basic time scale $\tau$, consists of $L$ Monte Carlo moves. At each such move, a lattice site is chosen randomly and updated according to the dynamical rules \eqref{model_stepping} to \eqref{model_attachment}. The simulations are performed long enough ($10^{10}$ steps) to ensure that the steady state of the system has been reached. The current and bound density are calculated by averaging over all the simulation steps and over all lattice sites.

In this article, we consider the traffic of two species of motors for long times, i.e.\ in the stationary state. Generally speaking, all the parameters for different motor species might be different. Here, as a simple starting point of theoretical work, we focus on the situation that the two motor species differ in only one parameter. We consider different unbinding probabilities, i.e.\ $\epsilon_1 \ne \epsilon_2$, in section \ref{Two species of motors with different unbinding probabilities} and different forward stepping probabilities, i.e.\ $\alpha_1 \ne \alpha_2$, in section \ref{Two species of motors with different stepping probabilities}.  We will not explicitly consider the case of different binding probabilities for different motor species, i.e.\ $\pi_{\rm ad,1} \ne \pi_{\rm ad,2}$, because this case is equivalent to that of different species with only different unbound densities, since the attachment probability is the product of binding probability for single motor $\pi_{\rm ad,i}$ and unbound motor density $\rho_{\rm ub,i}$, see the dynamical rule \eqref{model_attachment}.

Since we are interested in modeling molecular motor traffic, we choose the probabilities for motor movement as found experimentally for some specific molecular motors. We focus on the well-studied motor kinesin-1 for which all necessary parameters have been measured in single molecule experiments. Kinesin takes about 100 steps before unbinding from the filament \cite{Howard_2001}, so that we can estimate the ratio of stepping and unbinding probabilities to be $\alpha/\epsilon\approx 100$. Kinesin's microtubule desorption constant $K_{\rm D,MT}$ is of the order of $0.1-1\ \mu$M \cite{Gilbert_Johnson_1998, Seitz_Surrey_2006}, which fixes the ratio of the unbinding and binding probabilities $\epsilon/\pi_{\rm ad}=N_{\rm A}\ell^3K_{\rm D,MT}\approx 10^{-4}$ with the Avogadro constant $N_{\rm A}$, see Ref.\ \cite{Lipowsky_Mueller_2006}. We choose the discretization time unit $\tau$ so that the binding probability $\pi_{\rm ad}\equiv1$, which means that we have unbinding probability $\epsilon=10^{-4}$ and stepping probability $\alpha=10^{-2}$. Considering kinesin's velocity $v_0=\alpha\ell/\tau\approx 1\,\mu$m/s \cite{Howard_2001}, this means that our discretization time is $\tau\approx10^{-4}$\ s, which is sufficiently small to avoid discretization artifacts. The typical values of the probabilities used in this article are summarized in Table \ref{table_probability_value}.

\begin{table}
\begin{tabular}{c||c|c|c}
\hline
probability & stepping $\alpha$ & detachment $\epsilon$ & attachment $\pi_{\rm ad}$ \\ \hline
typical value& \multirow{2}{*}{$10^{-2}$} & \multirow{2}{*}{$10^{-4}$} & \multirow{2}{*}{1} \\
of kinesin & & & \\ \hline
\end{tabular}
\caption{Summary of the typical values of probabilities per time step $\tau=10^{-4}$\ s used in this article, which mimic the properties of the molecular motor kinesin-1.}
\label{table_probability_value}
\end{table}

\begin{figure*}
\centering
\includegraphics[width=16cm,clip]{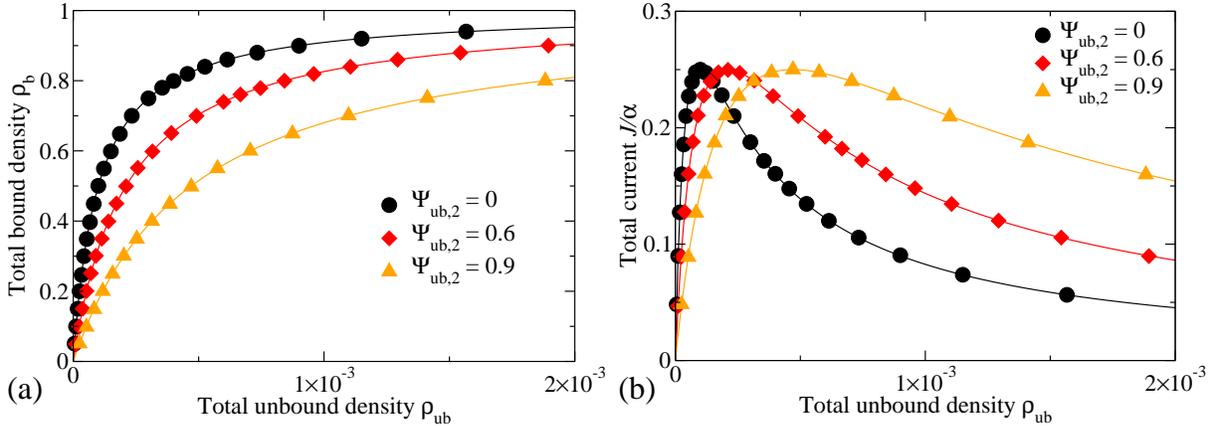}
\caption{(Color online) Traffic of two motor species with different unbinding probabilities. (a) Total bound density $\rho_{\rm b}$ and (b) normalized total current $J/\alpha$ (in the unit of $\tau^{-1}$) as a function of total unbound density $\rho_{\rm ub}$ for different fractions $\Psi_{\rm ub,2}$ of type-2 motors. The simulation results (data points) agree well with the mean field calculation of Eqs.\ \eqref{unbinding_bound_density}, \eqref{unbinding_effective_unbinding_probability} and \eqref{mean_field_current_density} (solid lines). The parameters are $\epsilon_1=10^{-4}$, $\epsilon_2=8\epsilon_1$, $\alpha=10^{-2}$, $\pi_{\rm ad}=1$ and $L=200$.}
\label{fig_different_unbinding}
\end{figure*}


\section{Two species of motors with different unbinding probabilities}
\label{Two species of motors with different unbinding probabilities}

In this section we consider the case of two species of motors which have different unbinding probabilities $\epsilon_1$ and $\epsilon_2$. The stepping probability $\alpha$ and the binding probability $\pi_{\rm ad}$ for both species are the same so that the indices of the corresponding symbols can be omitted.

Since the detachment and attachment rules \eqref{model_detachment} and \eqref{model_attachment} do not involve neighboring sites, there is no correlation of motors in the processes of binding and unbinding. Therefore the balances of attachment and detachment for both species of motors in the stationary state can be described by mean field theory as follows:
\begin{eqnarray}
\pi_{\rm ad} \rho_{\rm ub,1} (1-\rho_{\rm b})= \epsilon_1 \rho_{\rm b,1} ,
\label{unbinding_balance_of_type_1} \\
\pi_{\rm ad} \rho_{\rm ub,2} (1-\rho_{\rm b})= \epsilon_2 \rho_{\rm b,2} ,
\label{unbinding_balance_of_type_2}
\end{eqnarray} 
where $\rho_{\rm b,1}$ and $\rho_{\rm b,2}$ are the bound densities of type-1 and type-2 motors, respectively, and $\rho_{\rm b}=\rho_{\rm b,1}+\rho_{\rm b,2}$ is the total bound density. The terms on the left-hand side of Eqs.\ \eqref{unbinding_balance_of_type_1} and \eqref{unbinding_balance_of_type_2} represent the attachment of motors, with the mean field factor $(1-\rho_{\rm b})$ for the exclusion interaction between motors. The terms on the right-hand side represent the detachment of bound motors. Here we ignore the exclusion interaction when motors try to detach from the filament since the total unbound density of motors $\rho_{\rm ub}=\rho_{\rm ub,1}+\rho_{\rm ub,2}$ is very low, typically below $10^{-3}$, which corresponds to a bulk concentration of unbound motor proteins below $\rho_{\rm ub}/\ell^3 N_{\rm A}\approx1$\ M.

For convenience, we define the fractions
\begin{eqnarray}
\Psi_{\rm ub,i}\equiv\frac{\rho_{\rm ub,i}}{\rho_{\rm ub}}\quad\text{and}\quad \Psi_{\rm b,i}\equiv\frac{\rho_{\rm b,i}}{\rho_{\rm b}}
\end{eqnarray}
of the $i$-th type of motors in the unbound and bound states, respectively, with $\Psi_{\rm ub,1}+\Psi_{\rm ub,2}=\Psi_{\rm b,1}+\Psi_{\rm b,2}=1$. Because the two species of motors have different unbinding probabilities, the fractions of these two types of motors in the bound states are not equal to those in the unbound states, or more precisely, 
\begin{equation}
\frac{\Psi_{\rm b,1}}{\Psi_{\rm ub,1}}=\frac{\epsilon_2}{\epsilon_1}\frac{\Psi_{\rm b,2}}{\Psi_{\rm ub,2}} \ne \frac{\Psi_{\rm b,2}}{\Psi_{\rm ub,2}}.
\end{equation} 

From equations \eqref{unbinding_balance_of_type_1} and \eqref{unbinding_balance_of_type_2} we obtain the bound densities of both types of motors and the total bound density
\begin{equation}
\rho_{\rm b} = \frac{\pi_{\rm ad} \rho_{\rm ub}}{\pi_{\rm ad} \rho_{\rm ub}+\epsilon_{\rm eff}}
\label{unbinding_bound_density}
\end{equation}
with an effective unbinding probability
\begin{equation}
\epsilon_{\rm eff}=\epsilon_1 \Psi_{\rm b,1}+ \epsilon_2 \Psi_{\rm b,2}=\frac{\epsilon_1\epsilon_2}{\epsilon_1 \Psi_{\rm ub,2}+ \epsilon_2 \Psi_{\rm ub,1}}.
\label{unbinding_effective_unbinding_probability}
\end{equation}
Eq.\ \eqref{unbinding_bound_density} has the same form as the relation of binding and unbinding probabilities in the 1-species ASEP with attachment and detachment, except that the unbinding probability of the single motor species is replaced by the weighted average effective unbinding probability $\epsilon_{\rm eff}$. In the limit of $\Psi_{\rm ub,1}=0$ or $\Psi_{\rm ub,2}=0$, the effective unbinding probability becomes $\epsilon_{\rm eff}=\epsilon_2$ or $\epsilon_{\rm eff}=\epsilon_1$, respectively, and Eq.\ \eqref{unbinding_bound_density} reduces to the corresponding relation of the single-species ASEP. 

The total current $J$ of both species can be determined by mean field theory as the product of forward stepping probability $\alpha$, total bound density $\rho_{\rm b}$ of the two types of motors and density of vacancies $1-\rho_{\rm b}$,
\begin{equation}
J=\frac{\alpha}{\tau}\rho_{\rm b}(1-\rho_{\rm b}).
\label{mean_field_current_density}
\end{equation} 
The total motor current has a maximum at $\rho_{\rm b}=0.5$. It decreases for bound densities larger than $\rho_{\rm b}=0.5$, because overcrowding of bound motors leads to traffic jams.

Simulation results of the ASEP with two species of particles with different unbinding probabilities are in excellent agreement with mean field theory, see the example with $\epsilon_2=8\epsilon_1$ in Fig.\ \ref{fig_different_unbinding}. This observation strongly suggests that mean-field theory is exact in the case of two species of motors which differ only in the unbinding probabilities.

Results of the mean-field theory for the ASEP with two species of particles with different unbinding probabilities are in excellent agreement with simulation results (see the example with $\epsilon_2=8\epsilon_1$ in Fig.\ \ref{fig_different_unbinding}) and with the numerical solution of the associated stochastic process for small system sizes (not shown). This observation indicates that mean-field theory is exact in the case of two species of motors which
differ only in the unbinding probabilities.

As shown in Fig.\ \ref{fig_different_unbinding}, the total bound density $\rho_{\rm b}$ increases with the total unbound density $\rho_{\rm ub}$, as to be expected. The total current $J$ as function of unbound density $\rho_{\rm ub}$ has a maximum because overcrowding by motors causes traffic jams. The maximum shifts to higher unbound density with increasing fraction of type-2 motors, because type-2 motors have a larger unbinding probability than type-1 motors. Therefore a higher fraction of type-2 motors leads to smaller bound density and less traffic jam.


\section{Two species of motors with different stepping probabilities}
\label{Two species of motors with different stepping probabilities}

We next consider two motor species that differ only in the forward stepping probabilities, i.e.\ $\alpha_1\ne\alpha_2$. The binding probability $\pi_{\rm ad}$ and unbinding probability $\epsilon$ for these two species are the same so that the indices of the corresponding symbols can be omitted. Without loss of generality, we assume that type-1 motors have a larger stepping probability than type-2 motors. We will focus on the case that the second motor species has zero forward stepping probability $\alpha_2=0$, because we expect that this largest difference between $\alpha_1$ and $\alpha_2$ leads to the strongest effect. This choice is also motivated by \textit{in vitro} experiments, which addressed the traffic of mixtures of moving motors and non-moving mutant motors \cite{Seitz_Surrey_2006}.


\subsection{Traffic of two species of motors with different stepping probabilities}
\label{Traffic of two species of motors with different stepping probabilities}

As already discussed in section \ref{Two species of motors with different unbinding probabilities}, there are no correlations of motors in the processes of attachment and detachment. Therefore the balances of binding and unbinding processes for both kinds of motors can be described by mean field theory as given in Eqs.\ \eqref{unbinding_balance_of_type_1} and \eqref{unbinding_balance_of_type_2} with the simplified condition $\epsilon_1=\epsilon_2=\epsilon$. From these equations, the total bound density $\rho_{\rm b}$ of the two types of motors is given by
\begin{equation}
\rho_{\rm b}=\frac{\pi_{\rm ad}\rho_{\rm ub}}{\pi_{\rm ad}\rho_{\rm ub}+\epsilon}.
\label{stepping_bound_density}
\end{equation}
Eq.\ \eqref{stepping_bound_density} can be viewed as a special case of Eq.\ \eqref{unbinding_bound_density} with $\epsilon_1=\epsilon_2=\epsilon$, so that $\epsilon_{\rm eff}=\epsilon$, see Eq.\ \eqref{unbinding_effective_unbinding_probability}.

Since type-1 and type-2 motors have the same binding and unbinding probabilities $\pi_{\rm ad}$ and $\epsilon$, the fractions of both types of motors in the bound and unbound states are equal,
\begin{eqnarray}
\Psi_{\rm ub,i}=\Psi_{\rm b,i} \equiv \Psi_{\rm i}\quad\text{for }{\rm i}=1,2,
\end{eqnarray}
with $\Psi_{\rm 1}+\Psi_{\rm 2}=1$.

\begin{figure}
\centering
\includegraphics[width=8cm,clip]{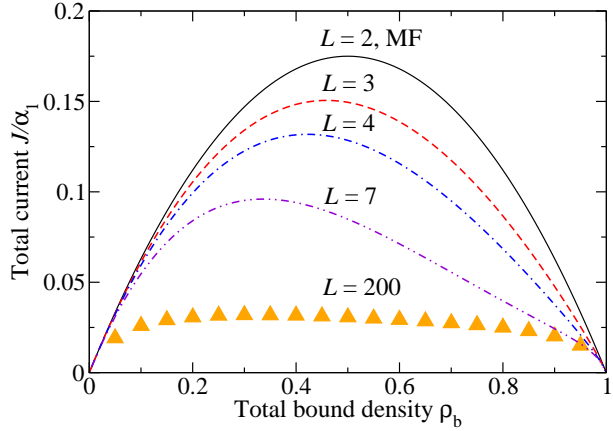}
\caption{(Color online) Traffic of two species of motors with different stepping probabilities. The normalized total motor current $J/\alpha_1$ (in the unit of $\tau^{-1}$) as a function of total bound density $\rho_{\rm b}$. Mean field theory as given by Eq.\ \eqref{stepping_current_mean_field} (solid black line) predicts a much larger total current than observed by simulation of a large system of size $L=200$ (data points). The other lines show the exact analytical results from solving Eq.\ \eqref{stepping_stationary_state} for small system sizes $L$ = 2, 3, 4, 7, with the result for $L=2$ being identical to the mean field result. The parameters are $\alpha_1=10^{-2}$, $\alpha_2=0$, $\pi_{\rm ad}=1$, $\epsilon=10^{-4}$ and $\Psi_2=0.3$.}
\label{fig_stepping_current_bound_density_small_system}
\end{figure}

The total current is given by the sum of the currents of the two motor species as $J=J_1+J_2$. Within the mean field approximation, the currents of the two motor species are
\begin{equation}
J_{\rm MF,i}=\frac{\alpha_{\rm i}}{\tau}\rho_{\rm b,i}(1-\rho_{\rm b})\quad\text{for }{\rm i}=1,2.
\end{equation}
Note that in this approximation, both species experience the exclusion interaction through the same hindrance factor $1-\rho_{\rm b}$, which depends on the total bound motor density $\rho_{\rm b}$. So the mean field total current is
\begin{equation}
J_{\rm MF}=J_{\rm MF,1}+J_{\rm MF,2}=\frac{\Psi_{\rm 1}\alpha_{\rm 1}+\Psi_{\rm 2}\alpha_{\rm 2}}{\tau}\rho_{\rm b}(1-\rho_{\rm b}).
\label{stepping_current_mean_field}
\end{equation}
As a function of the total bound density $\rho_{\rm b}$, $J_{\rm MF}$ has a maximum at $\rho_{\rm b}=0.5$ and is symmetric with respect to $\rho_{\rm b}=0.5$. However, simulations show that the actual value of the total motor current for large systems is much lower than the mean-field prediction, see Fig.\ \ref{fig_stepping_current_bound_density_small_system}. In addition, the total current as function of bound density is no longer symmetric with respect to $\rho_{\rm b}=0.5$ for $L>2$. These deviations from the mean-field prediction indicate strong correlations of bound motors which come from the blocking of type-1 motors by the non-moving type-2 motors.

To take these correlations into account, we consider the exact time evolution of the system. The possible configurations of the system can be labeled by the sequence of lattice site occupancies. For example, the configuration $\mathcal{C}=012$ of a system of size $L=3$ means that the first site is empty and the second and the third sites are occupied by a type-1 motor and a type-2 motor, respectively. At time $t$, the system is in  configuration $\mathcal{C}$ with probability $P(\mathcal{C}, t)$. At the next time step $t+\tau$, the system evolves stochastically following the rules \eqref{model_stepping}, \eqref{model_detachment} and \eqref{model_attachment}, and the probability $P(\mathcal{C}, t)$ to find the system in configuration $\mathcal{C}$ satisfies the Master equation
\begin{eqnarray}
P(\mathcal{C}, t+\tau)-P(\mathcal{C}, t)=
\sum_{\mathcal{C}^\prime}
M(\mathcal{C}, \mathcal{C}^\prime)P(\mathcal{C}^\prime, t).
\label{master_equation}
\end{eqnarray}
Here, the off-diagonal terms $M(\mathcal{C}, \mathcal{C}^\prime)$ of the transition matrix $\mathbf{M}$ with $\mathcal{C}^\prime \ne \mathcal{C}$ represent the probability of a transition from configuration $\mathcal{C}^\prime$ to configuration $\mathcal{C}$ during the time interval $\tau$, and the diagonal terms
\begin{equation}
M(\mathcal{C}, \mathcal{C})=-\sum_{\mathcal{C}^\prime \ne \mathcal{C}}
M(\mathcal{C}^\prime, \mathcal{C})
\label{master_equation_diagonal_term}
\end{equation}
represent the total exit probability from the corresponding configuration $\mathcal{C}$.

For long times, the system evolves into the stationary state with time-independent probabilities $P^{\rm st}(\mathcal{C})$, which satisfy
\begin{equation}
0=\sum_{\mathcal{C}^\prime}
M(\mathcal{C}, \mathcal{C}^\prime)P^{\rm st}(\mathcal{C}^\prime),
\label{stepping_stationary_state}
\end{equation}
and are therefore given by the null eigenvector of the the transition matrix $\mathbf{M}$. From the stationary distribution $P^{\rm st}$, it is straightforward to obtain the total bound density and total current by averaging over the appropriate configurations.

For a system of size $L$ with 2 motor species, there are in total $3^L$ possible configurations. Therefore the dimension of the transition matrix increases exponentially with the system size $L$ and Eq.\ \eqref{stepping_stationary_state} can in practice only be solved for very small systems.

The dimension of the transition matrix can be reduced based on the symmetry of the system. Since the system is homogeneous with periodic boundary conditions, configurations which differ by a translational shift are equivalent. For example, the configurations 102, 021 and 210 of a system of size $L=3$ are all equivalent and can be combined into one equivalence class. Since the translation operators form a cyclic group $\mathbb C_{\mit L}$, the number of equivalence classes for arbitrary size $L$ is equal to the cycle index $\mathcal{Z}(\mathbb{C}_L)$, which is given by Burnside's Lemma \cite{Harary_1994}:
\begin{equation}
\mathcal{Z}({\mathbb C_{\mit L}})=\frac{1}{L}\sum_{k=1}^{L}3^{\rm GCD\mit (k,L)}\approx \frac{1}{L}3^{L},
\label{cycle index}
\end{equation} 
where $\rm GCD\mit (k,L)$ is the greatest common divisor of $k$ and $L$ and the number 3 comes from the three possible configurations $0$, $1$ and $2$ for every filament site. This symmetry allows a reduction of the matrix size by combing all equivalent configurations into corresponding equivalence classes and appropriately modifying the transition probabilities, as described in detail in the appendix. In the limit of large $L$,  this procedure reduces the dimension $3^L$ of the transition matrix by a factor of $1/L$, as can be seen from Eq.\ \eqref{cycle index}.

\begin{figure}
\centering
\includegraphics[width=8cm,clip]{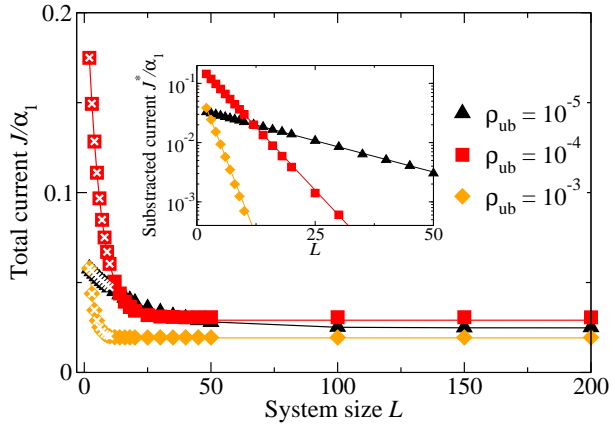}
\caption{(Color) Traffic of two species of motors with different stepping probabilities. The normalized total motor current $J/\alpha_1$ (in the unit of $\tau^{-1}$) decreases with the system size $L$ to a plateau value $J_{\infty}$. The simulation data (points) are well fitted by the exponential function of Eq.\ \eqref{stepping_current_large_system} (solid lines), which itself is a fit to the exact analytical solution of Eq.\ \eqref{stepping_stationary_state} for small systems (white x). The inset shows a semi-logarithmic plot of the normalized total current $J^*/\alpha_1=(J-J_{\infty})/\alpha_1$ (in the unit of $\tau^{-1}$) from which the plateau value has been subtracted. The parameters are $\alpha_1=10^{-2}$, $\alpha_2=0$, $\pi_{\rm ad}=1$, $\epsilon=10^{-4}$ and $\Psi_2=0.3$.}
\label{fig_stepping_finite_size}
\end{figure}

We then solve for the null eigenvector of the reduced transition matrix and obtain the exact total current for system of size $L=2$ to $L=10$. As shown in Fig.\ \ref {fig_stepping_current_bound_density_small_system}, the total current for system size $L=2$ is identical to the mean field total current \eqref{stepping_current_mean_field} as can be checked explicitly by exact solution of Eq.\ \eqref{stepping_stationary_state} for $L=2$. For systems with $L>2$, the particle-hole symmetry on the filament is lost, and the asymmetry of the total current as function of the total bound density with respect to $\rho_{\rm b}=0.5$ increases for increasing $L$, see Fig.\ \ref{fig_stepping_current_bound_density_small_system}.

As can also be seen in Fig.\ \ref{fig_stepping_current_bound_density_small_system}, as well as in Fig.\ \ref{fig_stepping_finite_size}, the total current decreases for increasing system size $L$. The reason is that in systems which are small compared to the motor run length, a stepping motor 'sees' the hole left by its image in front, which leads effectively to reduced congestion and larger total current. As the system size increases, this effect becomes less pronounced, so that the total current decreases and reaches a plateau value for system sizes comparable to the run length, which is 100 sites for our choice of parameters, see Fig.\ \ref{fig_stepping_finite_size}.

As shown in the inset of Fig.\ \ref{fig_stepping_finite_size}, a more detailed look at this finite-size effect reveals that the approach of the total current towards its plateau value $J_{\infty}$ can be approximated by the function
\begin{equation}
J(L)=J_{\infty}+ba^{-L}.
\label{stepping_current_large_system}
\end{equation}
Here, the two parameters $a$ and $b$, which characterize the exponential decay, depend on the dynamic parameters of the system such as forward stepping probability $\alpha_1$ or the fraction $\Psi_2$ of type-2 motors.

\begin{figure}
\centering
\includegraphics[width=8cm,clip]{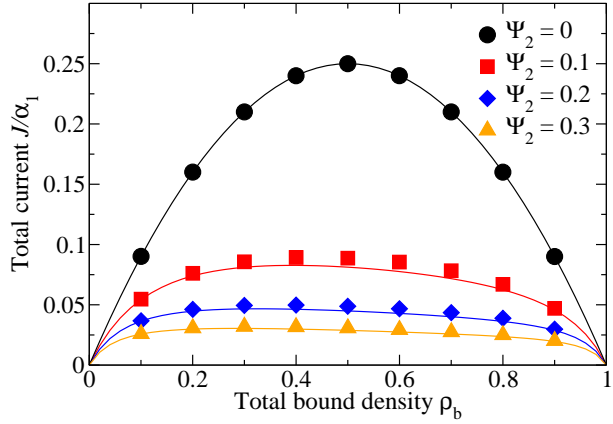}
\caption{(Color online) Traffic of two species of motors with different stepping probabilities. Normalized total motor current $J/\alpha_1$ (in the unit of $\tau^{-1}$) as a function of total bound density $\rho_{\rm b}$ for different fraction $\Psi_2$ of type-2 motors. The extrapolations (lines) following Eq.\ \eqref{stepping_current_large_system} from exact results for small systems of sizes $L=2-8$ are in fair agreement with the simulation results (data points). The parameters are $\alpha_1=10^{-2}$, $\alpha_2=0$, $\pi_{\rm ad}=1$, $\epsilon=10^{-4}$ and $L=200$.}
\label{fig_stepping_current_bound_density}
\end{figure}

We now use this finite-size effect to obtain the total current $J_{\infty}$ for large systems. We first obtain the exact results for the total current of small systems with size $L$=2 to 8 by solving Eq.\ \eqref{stepping_stationary_state}. Then we fit these total current values by Eq.\ \eqref{stepping_current_large_system} to estimate the plateau total current $J_{\infty}$. This extrapolation method gives fair agreement with the simulation results, see Figs.\ \ref{fig_stepping_finite_size} and \ref{fig_stepping_current_bound_density}.

Fig.\ \ref{fig_stepping_current_bound_density} shows the results for the total motor current as function of total bound density. Since non-moving type-2 motors tend to hinder the stepping events of type-1 motors and hence enhance the traffic jams, the current-density curves are asymmetric with respect to $\rho_{\rm b}=0.5$, with the position of the maximum shifted towards the low density region. As the fraction of type-2 motors increases, the total motor current decreases. This effect is quite dramatic: For instance, the total motor current drops more than one half as the fraction of type-2 motors increases from 0 to 0.1. We note that a similar effect has recently been discussed in the context of the traffic of RNA polymerases in transcription of bacterial ribosomal RNA \cite{Klumpp_Hwa2008}.

Our finding of the exponential finite-size scaling of the total current is different from the $1/L$-scaling observed in the single-species ASEP with periodic boundary conditions, which arises from the constraint of a fixed particle number in such a model \cite{Kanai_Tokihiro_2006}. The presence of attachment and detachment kinetics removes this constraint and leads to weaker finite-size effects. Accordingly, we do not observe any finite-size corrections for the case $\Psi_2=0$, which corresponds to a single species ASEP with periodic boundary conditions and particle attachment and detachment (data not shown). Exponential finite-size corrections to the total current, as observed for $\Psi_2>0$, are also obtained in single-species ASEPs with open boundaries in the low or high density phase \cite{Blythe_Evans_2007}. It appears that the exponential scaling can be obtained when there are interfaces between regions of different densities or boundary layers, i.e.\ interfaces at the boundary of the system. This is corroborated by the fact that the finite-size effect disappears for the single species ASEP with open boundary conditions with or without particle attachment and detachment kinetics, if the boundary rates are chosen so that there are no boundary layers. It is likely that the local traffic jams induced by the slow particles for $\Psi_2>0$ have an effect similar to that of jamming at interfaces between regions of different densities.


\subsection{Single moving motor with obstacle motors}
\label{Single moving motor with obstacle motors}

From a theoretical point of view, single tagged particles have been used to study interesting phenomena in ASEP-type models, such as shocks \cite{Mallick_1996}. 
The motion of a single labeled motor in the environment of many other motors has also been studied in single molecule experiments \cite{Seitz_Surrey_2006}.

These experiments motivate us to consider the case of only one moving type-1 motor which has a nonzero stepping probability $\alpha_1$ in the background of type-2 motors with stepping probability $\alpha_2=0$ which we will from now on call obstacle motors. Both types of motors have the same unbinding probability $\epsilon$ and binding probability $\pi_{\rm ad}$.

Because there is only a single type-1 moving motor, the total bound density given by Eq.\ \eqref{stepping_bound_density} is basically identical to the bound density of type-2 obstacle motors $\rho_{\rm b,2}$ provided that the system is sufficiently large.

\begin{figure}
\centering
\includegraphics[width=8cm,clip]{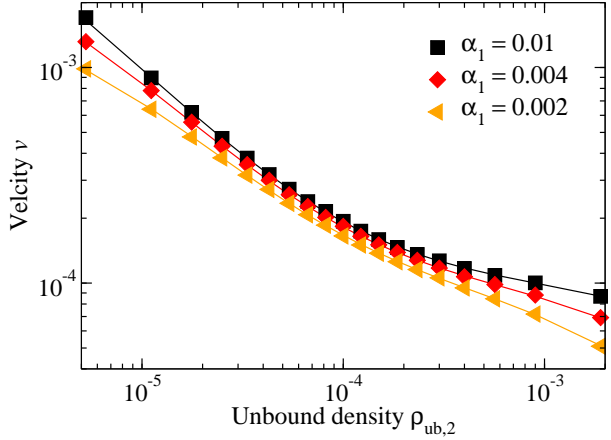}
\caption{(Color) Velocity $v$ (in the unit of $\ell/\tau$) of a single moving motor as a function of the unbound density $\rho_{\rm ub,2}$ of obstacle motors. The simulation results (data points) are in excellent agreement with the analytic results given by Eq.\ \eqref{stepping_single_moving_motor_velocity} (lines). The parameters are $\alpha_2=0$, $\pi_{\rm ad}=1$, $\epsilon=10^{-4}$ and $L=100$.}
\label{fig_stepping_single_moving_motor}
\end{figure}

Since we are interested in the movement of the bound type-1 motor, we will only take into account the bound state of the moving motor. By noticing that there is only one moving bound motor, we can label the states of the system by the site which is just in front of the moving bound motor. There are two states of this site: occupied by an obstacle motor or empty, and these states are denoted by $12$ and $10$, respectively.

The system changes from state $12$ to state $10$ with probability $\epsilon$ when the obstacle motor in front of the moving motor detaches from the filament:
\begin{eqnarray}
12& \xrightarrow{\ \epsilon\ } 10
\end{eqnarray}

Likewise, the state of the system changes from $10$ to $12$ with probability $\pi_{\rm ad} \rho_{\rm ub,2}$ when an obstacle motor binds to the empty site in front of the moving motor, and with probability $\alpha_{\rm 1} \rho_{\rm b,2}$ when the moving motor moves forward (with probability $\alpha_{\rm 1}$) and meets an obstacle motor (with probability $\rho_{\rm b,2}$): 
\begin{eqnarray*}
10&\xrightarrow{\pi_{\rm ad}\rho_{\rm ub,2}} &12 \ \ \text{ or } \ \ \underline{10}X \xrightarrow{\alpha_{\rm 1} \rho_{\rm b,2}}  0\underline{12}
\end{eqnarray*}
Here $X$ denotes a site that is occupied by an obstacle motor with probability $\rho_{\rm b,2}$ or free with probability $(1-\rho_{\rm b,2})$.

The transition matrix
\begin{equation}
\mathbf{S}= 
\left( \begin{array}{cc}
-\epsilon & \pi_{\rm ad} \rho_{\rm ub,2}+\alpha_{\rm 1} \rho_{\rm b,2} \\
\epsilon  & -\pi_{\rm ad} \rho_{\rm ub,2}-\alpha_{\rm 1} \rho_{\rm b,2} 
\end{array} \right)
\label{stepping_single_moving_motor_transition_matrix}
\end{equation}
for the state vector $\left(12, 10\right)^{\rm T}$ determines the stationary state of the system as the null eigenvector
\begin{equation}
E_0= 
\left( \frac{\pi_{\rm ad} \rho_{\rm ub,2}+\alpha_{\rm 1} \rho_{\rm b,2}}{\epsilon+ \pi_{\rm ad} \rho_{\rm ub,2}+\alpha_{\rm 1} \rho_{\rm b,2}}, \frac{\epsilon}{\epsilon+\pi_{\rm ad} \rho_{\rm ub,2}+\alpha_{\rm 1} \rho_{\rm b,2}} \right)^{\rm T}.
\label{stepping_single_moving_motor_stationary_state}
\end{equation}

Since only the state $10$ contributes to the motion of the moving bound motor, its velocity is given as
\begin{eqnarray}
v=\left( \begin{array}{cc}
0 , & \alpha_{\rm 1}\frac{\ell}{\tau} 
\end{array} \right)E_0 
=\frac{\alpha_{\rm 1}\epsilon}{\epsilon+\pi_{\rm ad}\rho_{\rm ub,2}+\frac{\alpha_{\rm 1}\pi_{\rm ad}\rho_{\rm ub,2}}{\epsilon+\pi_{\rm ad}\rho_{\rm ub,2}}}\frac{\ell}{\tau}.
\label{stepping_single_moving_motor_velocity}
\end{eqnarray}

As shown in Fig.\ \ref{fig_stepping_single_moving_motor}, the velocity of the moving bound motor decreases as the concentration of obstacle motors increases, and the exact Eq.\ \eqref{stepping_single_moving_motor_velocity} is in good agreement with the simulation data.


\section{Summary and discussion}
\label{Summary and discussion}

Motivated by cellular traffic, which is powered by several different species of molecular motors, we have studied the traffic of two motor species along the same track. For this purpose we have used a two-species ASEP model with periodic boundary conditions with attachment and detachment kinetics, in which the two motor species move into the same direction but have different velocities or unbinding rates. We have determined characteristic properties of the movements, such as motor density and total motor current, especially the current-density relationship, by computer simulations and by analytical calculations. As in the cell or in \textit{in vitro} experiments, our control parameter is provided by the densities of unbound motors.

We have considered two motor species that differ in only one property. If the two motor species only differ in the unbinding probabilities, increasing the fraction of motors with larger unbinding probability leads to a decrease of the total bound density of motors. As a consequence, upon increasing the fraction of motors with larger unbinding probability, the total current decreases in the low density region and increases in the high density region, see Fig.\ \ref{fig_different_unbinding}(b). For the case of two motor species with different velocities we have focused on the extreme case of non-zero and zero stepping probability for the two motor types. As to be expected, the total motor current decreases with increasing fraction of non-moving motors. Since  particle-hole symmetry is broken for two motor types with different stepping parameters, the current-density relationship becomes asymmetric with respect to the bound density $\rho_{\rm b}=0.5$. As shown in Fig.\ \ref{fig_stepping_current_bound_density_small_system}, its maximum shifts towards lower bound densities because non-moving motors hinder the stepping of moving motors and hence enhance traffic jams.

To obtain these results, we have employed computer simulations and analytical calculations. We have first used mean field theory to analyze our model. The attachment and detachment processes only involve single filament sites and therefore do not introduce new correlations into the model, so that mean field theory works well for the case of two species which differ only in their unbinding parameters. However, mean field theory fails to estimate the total  current of two motor species with different stepping probabilities, as illustrated in Fig.\ \ref{fig_stepping_current_bound_density_small_system}. In this case, the exclusion interaction leads to strong interference of the motors moving at different velocities and therefore to correlations and failure of mean field theory. In order to obtain analytical results in this latter case, we go beyond mean field and use finite-size scaling of the total current to extrapolate the total current value of small systems to the thermodynamic limit of large system size. This is possible because the total current decreases exponentially with the system size, as shown in Fig.\ \ref{fig_stepping_finite_size}.

This work presents a simple model for traffic by multiple motor species, which can be regarded as a starting point towards the understanding of intracellular transport with different kinds of molecular motors. Our model could be extended in many ways, such as by considering more than two different motor species which differ in multiple properties. Also, the model for the motors can be refined to include internal states of the stepping process or interactions between the motors which go beyond spatial exclusion. In the simple form presented here, our model gives a qualitative picture of the effects of traffic by several motor species. It makes predictions for the current-density relationship (Fig.\ \ref{fig_different_unbinding} and Fig.\ \ref{fig_stepping_current_bound_density}) and for the velocity of a single motor in the presence of immobile obstacles (Fig.\ \ref{fig_stepping_single_moving_motor}). It also suggests that the movements of different species of motors, and therefore of different kinds of cargo, can be cross-regulated, so that the movement of one type of motor is controlled indirectly by the presence or absence of other types of motors or by changes in the parameters of these motors. As an example, in the case of two species with different stepping probabilities, a small fraction of slow moving motors can strongly suppress the movement of fast moving motors and lead to a significant reduction of the current. This may be useful for the regulation of transport of different types of cargo along the same filament, both for \textit{in vivo} traffic and for \textit{in vitro} biomimetic applications.



\begin{acknowledgments}
SK was supported by Deutsche Forschungsgemeinschaft (Grants KL818/1-1 and 1-2) and by the National Science Foundation through the Center for Theoretical Biological Physics (Grant PHY-0822283).
\end{acknowledgments}


\appendix*
\section{}

In this appendix, we explain in detail how we calculate the exact total current of the system studied in section IV A. We illustrate the method for a system of size $L=2$. We recall that the type-1 motors have positive stepping probability $\alpha_1$ and type-2 motors have zero stepping probability $\alpha_2=0$. Both species have the same unbinding probability $\epsilon$ and the same binding probability $\pi_{\rm ad}$.

For a system with size $L=2$, there are $3^2=9$ configurations: 00, 01, 02, 10, 11, 12, 20, 21 and 22, where 0, 1 and 2 denote an empty site, a site occupied by a type-1 motor and a site occupied by a type-2 motor, respectively. We label these 9 configurations by $k=0,\ldots,8$ and combine them to the state vector $\left(00, 01, 02, 10, 11, 12, 20, 21, 22\right)^{\rm T}$. The corresponding transition matrix $\mathbf{M}$, as defined in Eqs.\ \eqref{master_equation} and \eqref{master_equation_diagonal_term}, is given by  
\begin{widetext}
\begin{eqnarray}
\mathbf{M}= 
\left(\begin{array}{ccccccccc} 
M_{0,0}  & \epsilon & \epsilon & \epsilon & 0        & 0        & \epsilon & 0        & 0\\
\pi_{\rm ad}\rho_{\rm ub,1} & M_{1,1}  & 0        & \alpha_1   & \epsilon & 0        & 0        & \epsilon & 0\\
\pi_{\rm ad}\rho_{\rm ub,2} & 0        & M_{2,2}  & 0        & 0        & \epsilon & \alpha_2 & 0        & \epsilon\\
\pi_{\rm ad}\rho_{\rm ub,1} & \alpha_1 & 0        & M_{3,3}  & \epsilon & \epsilon & 0        & 0        & 0\\
0        & \pi_{\rm ad}\rho_{\rm ub,1} & 0        & \pi_{\rm ad}\rho_{\rm ub,1} & M_{4,4}  & 0        & 0        & 0        & 0\\
0        & 0        & \pi_{\rm ad}\rho_{\rm ub,1} & \pi_{\rm ad}\rho_{\rm ub,2} & 0        & M_{5,5}  & 0        & 0        & 0\\
\pi_{\rm ad}\rho_{\rm ub,2} & 0        & \alpha_2 & 0        & 0        & 0        & M_{6,6}  & \epsilon & \epsilon\\
0        & \pi_{\rm ad}\rho_{\rm ub,2} & 0        & 0        & 0        & 0        & \pi_{\rm ad}\rho_{\rm ub,1} & M_{7,7}  & 0\\
0        & 0        & \pi_{\rm ad}\rho_{\rm ub,2} & 0        & 0        & 0        & \pi_{\rm ad}\rho_{\rm ub,2} & 0        & M_{8,8}\\
\end{array}  \right).
\end{eqnarray}
\end{widetext}
The diagonal terms $M_{k,k}$ are abbreviations for $M(\mathcal{C}^{(k)},\mathcal{C}^{(k)})$, $k=0,\ldots,8$, which represent the exit probability from configuration $\mathcal{C}^{(k)}$ during time interval $\tau$. In order to conserve the probability, the sum of $M_{k,k}$ and all other elements in the column $k$ is zero, compare Eq.\ \eqref{master_equation_diagonal_term}. For instance, $M_{0,0}=-2\pi_{\rm ad}\rho_{\rm ub,1}-2\pi_{\rm ad}\rho_{\rm ub,2}$ and $M_{1,1}=-\epsilon-\alpha_1-\pi_{\rm ad}\rho_{\rm ub,1}-\pi_{\rm ad}\rho_{\rm ub,2}$.

The stationary state of the system $P^{\rm st}$ is given by the null eigenvector of the transition matrix $\mathbf{M}$ as presented in Eq.\ \eqref{stepping_stationary_state}. By averaging all the contributions to the total current from the stationary probability distribution, we get the total current:
\begin{equation}
	J=\text{Tr}(\mathbf{D}\cdot P^{\rm st})/\tau L,
\end{equation}
with a diagonal matrix $\mathbf{D}$, which represents the contributions of each configuration to the total current. For $L=2$, this matrix is given by
\begin{equation}
	\mathbf{D}=\text{diag}\{0,\ \alpha_1,\ \alpha_2,\ \alpha_1,\ 0,\ 0,\ \alpha_2,\ 0,\ 0\}.
\end{equation}

The dimension of the transition matrix for the all $3^L$ configurations increases exponentially as the system size $L$ increases. With the help of translation symmetry, the dimension of transition matrix can be reduced by a factor $\sim 1/L$ as explained in Eq.\ \eqref{cycle index}, see also Table\ \ref{The translation symmetry reduces the size of the transition matrix}.

\begin{table}
\caption{The translation symmetry reduces the dimension of the transition matrix}
\begin{center}
\begin{tabular}{c||c|c}
\hline
 system size & dimension of original &  dimension of reduced \\
 $L$      & transition matrix        & transition matrix \\\hline
2	&	9	&	6	\\\hline
3	&	27	&	11	\\\hline
4	&	81	&	24	\\\hline
5	&	243	&	51	\\\hline
6	&	729	&	130	\\\hline
7	&	2187	&	315	\\\hline
8	&	6561	&	834	\\\hline
9	&	19683	&	2195	\\\hline
10	&	59049	&	5934	\\\hline
11	&	177147	&	16107	\\\hline
12	&	531441	&	44368	\\\hline
\end{tabular}
\end{center}
\label{The translation symmetry reduces the size of the transition matrix}
\end{table}

The translation operators form a cyclic group $\mathbb C_{\mit L}$. The quotient set of the cyclic group $\mathbb C_{\mit 2}$ contains $\mathcal{Z}(\mathbb{C}_2)=6$ equivalence classes, i.e. [00], [01], [02], [11], [12] and [22], which contain 1, 2, 2, 1, 2, 1 equivalent configurations, respectively. Each equivalence class represents all configurations which differ by a translational shift. For instance, the equivalence class [01] represents not only the configuration 01 but also the configuration 10 which can be translated into 01 by the cyclic group $\mathbb C_{\mit 2}$.

The corresponding reduced transition matrix is obtained by adding up all the transition probabilities within the same equivalence class. For $L=2$, the reduced transition matrix for the vector of equivalence classes $\left([00], [01], [02], [11], [12], [22]\right)^{\rm T}$ is
\begin{eqnarray}
&&\mathbf{M^{\prime}}=\\*
&&\left(\begin{array}{cccccc} 
M^{\prime}_{1,1} & 2\epsilon        & 2\epsilon        & 0                & 0         & 0 \\
2\pi_{\rm ad}\rho_{\rm ub,1}        & M^{\prime}_{2,2} & 0                & 2\epsilon        & 2\epsilon & 0 \\
2\pi_{\rm ad}\rho_{\rm ub,2}        & 0                & M^{\prime}_{3,3} & 0                & 2\epsilon & 2\epsilon \\
0                & 2\pi_{\rm ad}\rho_{\rm ub,1}        & 0                & M^{\prime}_{4,4} & 0         & 0 \\
0                & 2\pi_{\rm ad}\rho_{\rm ub,2}        & 2\pi_{\rm ad}\rho_{\rm ub,1}        & 0                & M^{\prime}_{5,5} & 0 \\
0                & 0                & 2\pi_{\rm ad}\rho_{\rm ub,2}        & 0                & 0 & M^{\prime}_{6,6} \\
\end{array}  \right) \nonumber,
\end{eqnarray}
with
\begin{eqnarray*}
M^{\prime}_{1,1}&=&-2\pi_{\rm ad}\rho_{\rm ub,1}-2\pi_{\rm ad}\rho_{\rm ub,2},\\
M^{\prime}_{2,2}&=&-2\epsilon-2\pi_{\rm ad}\rho_{\rm ub,1}-2\pi_{\rm ad}\rho_{\rm ub,2},\\
M^{\prime}_{3,3}&=&-2\epsilon-2\pi_{\rm ad}\rho_{\rm ub,1}-2\pi_{\rm ad}\rho_{\rm ub,2},\\
M^{\prime}_{4,4}&=&-2\epsilon,\\
M^{\prime}_{5,5}&=&-4\epsilon,\\
M^{\prime}_{6,6}&=&-2\epsilon.
\end{eqnarray*}
The null eigenvector of the reduced transition matrix $\mathbf{M^{\prime}}$ gives the stationary probability distribution of equivalence classes $P^{\prime {\rm st}}$. 
The numbers of equivalent configurations for the equivalence classes $\left([00], [01], [02], [11], [12], [22]\right)$ are summarized as a weight vector:
\begin{equation}
	W=\left(1,\ 2,\ 2,\ 1,\ 2,\ 1\right)^{\rm T},
\end{equation}
and the stationary probability distribution of the equivalence classes $P^{\prime {\rm st}}$ satisfies the normalization condition:
\begin{equation}
	W^{\rm T}\cdot P^{\prime {\rm st}}=1.
\end{equation}
The contributions of the classes $\left([00], [01], [02], [11], [12], [22]\right)^{\rm T}$ to the total current are described by
\begin{equation}
\mathbf{D}^{\prime}=\text{diag}\{0,\ \alpha_1,\ \alpha_2,\ 0,\ 0,\ 0\}.
\end{equation}
Finally, the total current is obtained by averaging the stationary probability distribution of equivalence classes, weighted by the number of equivalent configurations within the classes:
\begin{equation}
J=(W^{\rm T}\cdot\mathbf{D}^{\prime}\cdot P^{\prime {\rm st}})/\tau L.
\end{equation}



\addcontentsline{toc}{section}{References}

\end{document}